\documentclass[12pt,epsfig]{article}
\newcommand{\ba}{\begin{array}{c}}
\newcommand{\baz}{\begin{array}{cc}}
\newcommand{\bad}{\begin{array}{ccc}}
\newcommand{\bav}{\begin{array}{cccc}}
\newcommand{\ea}{\end{array}}

\usepackage{amssymb} 
\usepackage{epsfig}
\usepackage{float}
\newcommand{\be}{\begin{equation}}
\newcommand{\ee}{\end{equation}}
\newcommand{\bea}{\begin{eqnarray}}
\newcommand{\eea}{\end{eqnarray}}
\begin{document}

\begin{center}
\bf {Invariants of Lepton Mass Matrices and \\
CP and T Violation in Neutrino Oscillations}
\end{center}

\begin{center}
C. Jarlskog 
\end{center}

\begin{center}
{\em Division of Mathematical Physics\\
LTH, Lund University\\
Box 118, S-22100 Lund, Sweden}
\end{center}

\begin{abstract}

CP and T asymmetries in neutrino oscillations, in vacuum as
well as in matter, are expressed in terms 
of invariant functions of lepton mass matrices.

\end{abstract}

\section{Introduction}

Four decades have passed since the unexpected discovery \cite{cp-discovery} of CP violation,
originally seen in two decay modes of
$K_L$ . For about 25 years the superweak ansats
by Wolfenstein \cite{lincoln64} accounted for all observed effects, including
CP violation in the semileptonic decay modes of $K_L$, and provided a simple 
and intuitive explanation for
why CP violation was not seen elsewhere in particle physics.
Nowadays the standard framework for understanding 
CP violation is the 
electroweak model \cite{sm} (hereafter referred to as the standard model)
and the Kobayashi-Maskawa scheme \cite{km} within it.
Indeed the remarkable experimental determination of the quantity $\epsilon^\prime / \epsilon$
in $K_L$ decays 
and the discovery of CP violation in the decays of B-mesons, at SLAC and KEK are in agreement
with the predictions 
of the standard model with three families. Nonetheless, understanding CP
violation still remains a great challenge, within a broad area in physics. 
One faces deep questions such as what is the 
source of CP violation that goes into generating
the baryon asymmetry of the universe and why is the
theta parameter of QCD so small that it has not been seen. 
There are also some perhap simpler 
questions, one being: is there CP violation in the leptonic sector?
The minimal standard model gives a no as the answer to the latter question
because it assumes that the neutrinos are massless. But 
nowadays there 
is evidence for neutrino oscillations, a phenomenon that requires massive neutrinos \cite{bruno}.
Therefore the minimal standard model needs to be modified, but we don't know how.
In this article, I shall address the issue of CP violation
in neutrino oscillations using the method of invariant functions of mass matrices
(\cite{ceja85a},\cite{ceja87}). In this article, first a
 short introduction to this
method is presented in the next section followed by application to 
neutrino oscillations, in vacuum as well as in matter, in the following sections.    

\section{Invariant functions of mass matrices}
\label{sec-quarks}

Consider first the quark sector of the standard model with three families. 
The identity of the
quarks is encoded in the three-by-three quark mass matrices
$M_u$ and $M_d$, 
for the up-type and down-type quarks respectively.
However, these mass matrices are basis dependent. Given any pair
$M_u,~M_d$ one can obtain other pairs through 
unitary rotations, as will be described below, without affecting
the physics. The measurable quantities must be basis independent and
therefore they are "invariant functions" under
such rotations. These functions were introduced in \cite{ceja85a} and studied in 
detail in \cite{ceja87}. 
Actually, what enters, in the standard model, is the pair
\begin{equation} 
S_u \equiv M_u M_u^\dagger~,~S_d \equiv M_d M_d^\dagger 
\end{equation}
\noindent For
simplicity, we shall refer to these quantities as 
mass matrices. This should cause no confusion because
the underlying mass matrices, $M_u$ and $M_d$,
do not enter in what follows.  

An invariant
function $f(S_u, S_d)$ is a function that does not change under the transformation
\begin{equation}
S_u \rightarrow X S_u X^\dagger,~S_d \rightarrow X S_d X^\dagger
\end{equation}
\noindent where $X$ is an arbitary three-by-three unitary matrix. 
Evidently, the traces of powers of the above quantities,
$tr(S_u^k ~S_d^l)$, are such invariant functions. Note that the corresponding determinants 
are not independent
invariant functions because any determinant can be expressed 
as a function of traces. For a detailed
discussion of these invariants see \cite{ceja87}. 
        
When dealing with CP violation in the standard model with three families, 
a central role is played by the commutator of the quark 
mass matrices, $\left[ S_u,S_d \right] $ (see (\cite{ceja85a}, \cite{ceja85}))
The determinant of this commutator is an invariant function of 
mass matrices given by (\cite{ceja85}, \cite{ceja85a})
\begin{equation}
det \left[ S_u,S_d \right]  = 2i J~ v(S_u) v(S_d)
\label{detquark}
\end{equation} 
\noindent where $J$ is the CP-invariant of the quark mixing matrix $V$,
\begin{equation}
Im (V_{\alpha j}V_{\beta k}V^\star_{\alpha k}V^\star_{\beta j})= J~\sum_{\gamma , l}^{}
\epsilon_{\alpha\beta\gamma} \epsilon_{jkl}
\end{equation}
\noindent J is equal \cite{stora88} to twice the area of 
any of the six by now well-known unitarity triangles. The quantities 
$v(S_u) $ and $v(S_d) $ are 
Vandermonde determinants as follows. Denoting the three eigenvalues of $S_u$ by
$x_i$, $x_1=m^2_u, ~x_2=m^2_c, ~x_3=m^2_t$, we have
\begin{equation}
v(S_u) = \sum_{i,j,k}^{} \epsilon_{ijk} x_j x^2_k =
(m^2_u-m^2_c)(m^2_c -m^2_t)(m^2_t -m^2_u)  
\end{equation}
\noindent and similarly $v(S_d)= (m^2_d-m^2_s)(m^2_s -m^2_b)(m^2_b -m^2_d).$ 

In the standard model with three families, the nonvanishing of the above determinant gives 
the if and only if condition for 
CP violation in the quark sector. In fact it manifestly unifies the 14 conditions needed,
such as the condition
that no two quarks with the same charge are allowed 
to be degenerate if CP is to be violated, 
or the conditions
that none of the mixing angles nor the phase 
angle is allowed to assume its maximal or mininmal value.
In other words, the commutator automatically 
keeps track of these requirements and is thus useful for checking
whether a specific model violates CP or 
not (for a pedagogical discussion see \cite{cpboken}). 

The determinant in Eq.(\ref{detquark}) appears naturally in computations involving CP
violation when all the six quarks enter on equal footing. 
Examples are the renormalization
of the $\theta$-parameter of QCD by the electroweak interactions and the calculation of the
baryon asymmetry of the universe in the standard model. The determinant, 
in a truncated form, enters in many more
computations such as when computing the electric dipole moment of a quark, say the
down quark. Since in such a computation, the down quark appears in the external legs, 
it is tacitly assumed
that we know the identity of this quark, i.e., $m_d \neq m_s$ and $m_d \neq m_b$. 
Therefore the factors
$(m^2_d-m^2_s)$ and $(m^2_b -m^2_d)$ will be missing but all 
the other factors will be present. 

The above commutator is the simplest in a family of comutators of 
functions of mass matrices \cite{ceja85a},
\begin{equation}
C(f,g) \equiv \left[ f(S_u),g(S_d ) \right]
\label{commutator}
\end{equation}
\noindent $f$ and $g$ being functions that are diagonalised with 
the same unitary matrices that
diagonalise $S_u$ and $S_d$ respectively. The determinants of these
commutators, which are also invariant functions, are given by 
\begin{equation}
det \left[ f,g \right] = 2i J~ v(f) v(g) 
\label{fgdet}
\end{equation}
\noindent where $v(f) \equiv v(f(S_u))$, and $v(g) \equiv v(g(S_d))$. 
More explicitly
\begin{equation}
v(f) = \sum_{i,j,k}^{} \epsilon_{ijk} f_j f^2_k~= 
(f_1-f_2)(f_2-f_3)(f_3-f_1)
\label{vander}
\end{equation} 
\noindent The $f_j$ denote the three eigenvalues of the matrix 
$f(S_u)$ and the quantities 
related to the down-type quarks are defined similarly.

An essential point is that Eq.(\ref{fgdet}) holds irrespectively of whether $f$ and $g$ 
are hermitian or not. This property makes the above formalism applicable to 
neutrino oscillations, as we shall see here below. 

\section{Neutrino oscillations in vacuum}
 
With massive neutrinos, CP violation in neutrino oscillations could manifest itself by the 
difference of the rates of reactions
$\nu_\alpha \rightarrow \nu_\beta$ and $\bar{\nu}_\alpha \rightarrow \bar{\nu}_\beta $.
Here $\alpha$ and $\beta$ ($\alpha \neq \beta$) stand for $e, \mu, \tau$. 
The difference of these
rates is found \cite{barger} to be proportional to 
\begin{equation}
sin(2\phi_{12})+ sin(2\phi_{23}) +sin(2\phi_{31}) = -4sin(\phi_{12}) sin(\phi_{23}) sin(\phi_{31})
\end{equation}
\begin{equation}
\phi_{jk}= (m^2_j-m^2_k) \xi,~~~  \xi = {L \over 4E} 
\label{defphi}
\end{equation}  
\noindent $m$'s being the neutrino masses;
E denotes the
neutrino energy and L is the distance from the source
(for a recent review see \cite{bilenky04}). Because of CPT invariance, the 
above combination of $sin$'s also appears when testing 
time-reversal asymmetry by comparing the rates of
$\nu_\alpha \rightarrow \nu_\beta$ and $\nu_\beta \rightarrow \nu_\alpha$. 
 
As was noted before \cite{ceja87}, in models with quark-lepton universality the results 
for the quark sector, presented in the previous section, can be 
extended to the leptonic sector by trivial substitutions, $M_u \rightarrow M_\nu$ and
$M_d \rightarrow M_l$
where $M_\nu$ and $M_l$ denote the
neutrino  and the charged lepton mass matrices. 
An interesting question is then: how is the sum (or product) of the above three $sin$'s 
related to the commutator of lepton mass matrices?
One would expect
that there be a relationship because, as far as neutrino oscillations are concerned, the leptonic 
sector is
essentially a copy of the quark sector. An essential point
is that the possible Majorana nature of neutrinos
is known not to be important in the computation of the rates (see \cite{petcov87}, 
\cite{langacker87}).
The relevant
formulae are obtained by assuming that there are three active neutrinos and
that their "effective" mass matrix is three by three. 
This pattern emerges in many models, some based on the see-saw
mechanism (for a recent review see \cite{ramond} ) as well as in
models \`{a} la Weinberg \cite{weinberg79} where the effective operator that generates 
the neutrino masses 
involves only the three active left-handed neutrinos.  

To answer the question concerning the relationship
between the above sum or product of the three $sin$'s and the lepton mass matrices, we
introduce, in analogy with the case of the quarks,
\begin{equation}
S_\nu \equiv M_\nu M_\nu^\dagger~,~S_l \equiv M_l M_l^\dagger
\end{equation}
\noindent where $M_\nu$ and $M_l$ are the three-by-three neutrino and charged lepton mass matrices 
respectively. We introduce the commutators
\begin{equation} 
\Delta^\pm \equiv \left[ e^{\pm 2i\xi S_\nu}~,~ S_l
\right]
\label{vaccom}
\end{equation}  
\noindent where the unitary matrices $U^\pm \equiv e^{\pm 2i\xi S_\nu}$
are inverses of one another and $\Delta^+ = (\Delta^-)^\dagger$.
The determinant of these commutators are invariant functions of
lepton mass matrices. Using Eq.(\ref{fgdet}) we have  
\begin{equation}
det\Delta^\pm= 2i~J_{\nu}~v(S_l)v(e^{\pm 2i\xi S_\nu })
\label{vacdet}
\end{equation} 
\noindent Here $J_{\nu}$ is the leptonic analogue of the CP invariant of the quark mixing matrix. 
Here it is more convenient to use the index $\nu$ instead of "lep" (for leptons) 
because when dealing with oscillations in matter, in the next section, the notation is
easily generalised. $J^\prime_{\nu}$ and $J^\prime_{{\bar \nu}}$ will then denote
the corresponding quantities for neutrino and antineutrino oscillations in matter.    

Furthermore, just as in the case of the quarks, $J_{\nu}$ is simply twice the area of 
any of the six leptonic unitarity
triangles. The two Vandermonde determinants in the above equation are given by
\begin{equation}
v(S_l)= (m^2_e-m^2_\mu)(m^2_\mu -m^2_\tau)(m^2_\tau -m^2_e)
\label{defsl}
\end{equation}
and
\begin{equation}
v(e^{^\pm 2i\xi S_\nu }) =  \mp 8i e^{\pm 2i\xi trS_\nu }sin(\phi_{12}) 
sin(\phi_{23}) sin(\phi_{31})
\end{equation}
\noindent where $trS_\nu = (m^2_1 + m^2_2 + m^2_3)$ and the $m$'s are again the
neutrino masses. The exponential factors are the determinants of $U^\pm$.
Putting these results together, we find
\begin{equation}
det\Delta^\pm = \pm 16 v(S_l) \left[ J_{\nu} sin(\phi_{12}) 
sin(\phi_{23}) sin(\phi_{31})\right] 
e^{\pm 2i\xi tr S_\nu}
\label{detavac}
\end{equation}
Note that the phase of the determinant is determined by the sum of the
neutrino masses. Evidently  
any of the relations $det\Delta^{\pm} \neq 0$ 
provides the necessary and sufficient condition for having CP violation   
in reactions $\nu_\alpha \rightarrow \nu_\beta$ and 
$\bar{\nu}_\alpha \rightarrow \bar{\nu}_\beta$.
The presence of the factors involving
the masses of charged leptons is essential. These keep track of the identity of neutrinos. 
For example, for $m_e = m_\mu$ the electron and muon neutrinos would be indistinguishable and
therefore there would be no CP violation. $det\Delta^\pm \neq 0$ also provides the if 
and only if condition for T-violation when comparing the reactions
$\nu_\alpha \rightarrow \nu_\beta$ and $\nu_\beta \rightarrow \nu_\alpha$. 

\section{Neutrino oscillations in matter}

At a first glance, the formalism for CP and time-reversal violation in
neutrino oscillations in matter looks deceptively similar to that in vacuum. 
It would seem that all one needs to do is to distinguish the masses and
mixings in matter by simply putting say primes on corresponding
quantities in vacuum, as we shall do here below,
and by introducing explicit indices $\nu$ and $\bar{\nu}$
to keep track of whether we are dealing with neutrino or antineutrino
propagation in matter. 
However, there are subtle points to be taken into account, as we shall see soon.
 
Consider first the case of neutrinos. To the leading order, in a frame where the charged 
lepton mass matrix is diagonal, 
the neutrino mass matrix, $S_\nu = M_\nu M_\nu^\dagger$, is replaced by  
an effective mass matrix in matter,
$S^\prime_\nu = M_\nu^\prime M_\nu^{\prime \dagger}$, given by
(\cite{lincoln78}, \cite{smirnov85})
\begin{equation}
S^\prime_\nu= S_\nu  + {\bf \Pi}, ~~~~{\bf \Pi} 
~\equiv 
diag (\rho, 0, 0)
\label{numatter}
\end{equation}
\noindent where $\rho$ is proportional to the neutrino momentum as well as 
the density of the electrons in matter 
(for a review see, for example, \cite{bilenky99}). 
Here, we shall treat $\rho$ as a constant. The reality is
generally far more complicated. But here we are primarily interested in
exploring the structure of matter oscillations rather than making
realistic calculations. Thus we construct 
the matter analogue of any of the two vacuum commutators and take its determinant. We 
take $\Delta^+$, drop the superscript and define 
\begin{equation}
\Delta^\prime_\nu \equiv \left[ e^{2i\xi S^\prime_\nu},~ S_l
  \right]
\label{deltaprime}
\end{equation}  
From Eq.(\ref{vacdet}) we have  
\begin{equation}
det\Delta_\nu^\prime=16\left[ J^\prime_\nu~v(S_l)sin(\phi^\prime_{12}) 
sin(\phi^\prime_{23}) sin(\phi^\prime_{31})\right] 
e^{2i\xi tr S^\prime_\nu } 
\label{detprime}
\end{equation}
Here the primed quantities are the matter analogues of the unprimed ones in
the previous section and, as before, the phase of this determinant is fixed by the sum of the 
neutrino masses, now taken in matter. Thus
$trS^\prime_\nu =(m^{\prime 2}_1 + m^{\prime 2}_2 + m^{\prime 2}_3)$; 
$m^\prime$'s  being the neutrino masses in matter, etc. 
The subscript $\nu$ is a reminder that we are
dealing with neutrinos propagating in matter. Evidently $J^\prime_\nu \rightarrow J_{\nu}$ as 
$\rho \rightarrow 0$.
 
Actually, this determinant is {\it not} what enters when testing CP violation in 
neutrino oscillations in matter through the rates discussed above. 
It will, however, enter if we were to do the science-fictional experiment of comparing
the rates of $\nu_\alpha \rightarrow \nu_\beta$ in matter with that of 
$\bar{\nu}_\alpha \rightarrow \bar{\nu}_\beta $ in the corresponding {\it antimatter}.
It is ironic that the very same baryon asymmetry of the universe, that owes its existence
to CP violation, forbids us to test CP violation in matter in the above "straight-forward" fashion.

Turning now to antineutrino oscillations in matter, the effective mass matrix,
$S^\prime_{\bar{\nu}} = M_{\bar{\nu}}^\prime M_{\bar{\nu}}^{\prime \dagger}$,
to the leading order is given by   
\begin{equation} 
S^\prime_{\bar{\nu}} = S_\nu  - {\bf \Pi }= S_\nu - diag (\rho, 0, 0)
\label{antinumatter}
\end{equation}
\noindent Because of the different sign of the added term for neutrinos and antineutrinos,
the effective neutrino mass of the j-th neutrino,
$m^\prime_j$, is in general not equal to that of the corresponding antineutrino,
$\bar{m}^\prime_j$. Nor are the mixing angles and thus
the corresponding $J$'s in general the same. 
This is the reason why we need to introduce not only primes to 
indicate the presence of matter but also 
appropriate $\nu$ and $\bar{\nu}$ indices to keep track of whether we are talking
about neutrinos or antineutrinos. Defining the commutator relevant to antineutrinos, we have
\begin{equation}
det\Delta_{\bar{\nu}}^\prime=16 v(S_l) \left[ J^\prime_{\bar{\nu}} 
sin({\bar{\phi}}_{12}^\prime) 
sin({\bar{\phi}}_{23}^\prime) sin({\bar{\phi}}_{31}^\prime) \right] 
 e^{2i\xi tr S^\prime_{\bar{\nu}}}
 \end{equation}
Depending on the density profile of matter, the above expressions could, however, 
be relevant when testing time-reversal violation. The 
quantity $det\Delta_\nu^\prime$ would enter when comparing the
rates of $\nu_\alpha \rightarrow \nu_\beta$ and $\nu_\beta \rightarrow \nu_\alpha$ in matter 
and $det\Delta_{\bar{\nu}}^\prime$ for corresponding antineutrinos processes again in matter.
 
Since the matter contribution to the mass matrices,
$\bf \Pi$, commutes with the lepton mass matrix
$S_l$ (which is diagonal), we have 
\begin{equation}
\left[S_\nu, S_l \right] = 
\left[S^\prime_\nu, S_l \right] = \left[S^\prime_{\bar{\nu}}, S_l \right]
\end{equation}
Taking the determinants of these matrices and removing the common factor involving
the charged leptons, gives  
\begin{equation}
J_{\nu}~v(S_\nu) = J^\prime_\nu ~v(S^\prime_\nu) = J_{\bar{\nu}}^\prime ~
v(S^\prime_{\bar{\nu}})
\label{jvequality}
\end{equation}
This result can be found in Refs.\cite{naumov92} and \cite{bill00}.
\noindent Again $v(S_\nu) = (m^2_1- m^2_2)(m^2_2 -m^2_3)(m^2_3 -m^2_1)$, the 
$m_j$'s being the neutrino masses in vacuum; $v(S^\prime_\nu)$ 
and $v(S^\prime_{\bar{\nu}})$ are   
the corresponding expressions with neutrino and antineutrino 
masses in matter respectively.
A more detailed discussion of some further matter effects 
is given in Refs. \cite{bill02} and \cite{bill03}. 

We shall now consider the next simplest commutator in the series, i.e.,
$\left[S^{\prime 2}_\nu,~ S_l \right]$.
The determinant of this commutator can be written down immediately, in
terms of the matter quantities, using the general formula Eq.(\ref{fgdet}). 
We have
\begin{equation}
det \left[S^{\prime 2}_\nu, S_l \right] = 2 i J^\prime_\nu ~v(S_l)
v(S^{\prime 2}_\nu) 
\label{dets2}
\end{equation}
\noindent where 
\begin{equation}
 v(S^{\prime 2}_\nu)= w(S^\prime_\nu)v(S^\prime_\nu)
\end{equation}
\begin{equation}
w(S^\prime_\nu) = (m^{\prime 2}_1 + m^{\prime 2}_2)
(m^{\prime 2}_2 + m^{\prime 2}_3) (m^{\prime 2}_3 + m^{\prime 2}_1)=
{1 \over 3} \{ (tr S^\prime_\nu )^3 - tr S^{\prime 3 }_\nu \}
\label{defw}
\end{equation}  
The determinant in Eq.(\ref{dets2}= can also be computed directly, in terms of
vacuum quantities, by substituting
$S^\prime_\nu = S_\nu  + \bf \Pi$. A straight-forward
calculation gives
\begin{equation}
det \left[S^{\prime 2}_\nu,~ S_l \right] = 2 i J_{\nu} ~ v(S_l)
v(S_\nu) \left[ w(S_\nu) + \rho I_1 + \rho^2 I_2 \right]
\label{directs2}
\end{equation}
\noindent where
\[ I_1 = (trS_\nu)^2 - \sum_\alpha^{} \mid V_{\alpha 1} {\mid}^2 m^4_\alpha
\]
\[I_2 = trS_\nu - \sum_\alpha^{} \mid V_{\alpha 1} {\mid}^2 m^2_\alpha
\]
\noindent Here all the masses and mixings refer to vacuum quantities.
Note that the coefficient of $\rho^3$ vanishes because
the ${\bf \Pi } \left[ S^n, S_l  \right] {\bf \Pi } =0.$ 
Equating the two expressions for the determinant and using that
$J^\prime_\nu ~v(S^\prime_\nu) = J_{\nu} ~v(S_\nu)$ gives
\[w(S^\prime_\nu) = w(S_\nu) + \rho I_1 + \rho^2 I_2\] 
Rewritting this relation in terms of traces gives
\[ (tr S^\prime_\nu )^3 - tr S^{\prime 3 }_\nu = (tr S_\nu )^3 - tr S^3_\nu
+ 3 \rho (tr S_\nu )^2 - 3 tr (S^2_\nu {\bf \Pi }) + 3 \rho^2 tr S_\nu
-3 tr (S_\nu {\bf \Pi}^2) 
\]
This relation can easily be checked by computing its left-hand side,
in terms of vacuum quantities, using again
$S^\prime_\nu = S_\nu  + {\bf \Pi}$. 

For antineutrinos the corresponding results are easily obtained by flipping the 
sign of ${\bf \Pi }$ and therefore also that of $\rho$.

\section{Oscillations in matter with low density}

The results obtained in the previous section hold to all orders in the
matter-related parameter $\rho$. Here we would like to examine
how the results look like in the low density limit, i.e., when
terms of order $\rho^2$ and higher can be neglected. We shall consider the case
of neutrinos, extension to antineutrinos being trivial. Beginning 
with the matter commutator in Eq.(\ref{deltaprime}) we write
\begin{equation}
det\Delta_\nu^\prime= det\Delta_\nu + \rho R + {\it O}(\rho^2)
\label{orderrho} 
\end{equation}
where the first term in the RHS is the
vacuum contribution. To compute $R$ we expand the exponential in Eq.(\ref{deltaprime}) using
Eq.(\ref{numatter}) and find
\begin{equation} 
\Delta_\nu^\prime= \Delta_\nu +
\sum_{n=1}^{\infty} {(2i\xi)^n \over n! }
\sum_{k=0}^{n-1} \left[ S_\nu^{n-1-k} {\bf \Pi }
S_\nu^k, S_l \right] +{\it O}(\rho^2)
\end{equation}
Taking the determinant of the RHS we find
\begin{equation}
R = -v(S_l)\sum_{\alpha, \beta, \gamma, \sigma}^{}E(\alpha, \beta, \gamma, \sigma)
F(\alpha, \beta, \gamma, \sigma)
\label{defr}
\end{equation}
where
\begin{equation}
E(\alpha, \beta, \gamma, \sigma)=\sum_{rst}^{}\epsilon_{rst}
(\alpha 1)^\star(\alpha r)
(\beta r)^\star(\beta s)(\gamma s)^\star(\gamma t)(\sigma t)^\star(\sigma 1)
\label{defe}
\end{equation}
\begin{equation}
F(\alpha, \beta, \gamma, \sigma)= 
({e^{2i\xi m^2_\alpha} - e^{2i\xi m^2_\sigma} \over m^2_\alpha - m^2_\sigma })
e^{2i\xi (m^2_\beta +m^2_\gamma )}
\label{deff}
\end{equation}
Here, for simplicity, we have introduced the short-hand notation
$(\alpha r) \equiv V_{\alpha r}$, $V$ being the vacuum lepton mixing matrix. 
Note that all the dummy indices in the above sums run from one to three.
Note that $E^\star (\alpha, \beta, \gamma, \sigma) = -E(\sigma, \gamma, \beta, \alpha)$
and since $F$ is symmetric under $\beta \leftrightarrow \gamma$ as 
well as under $\alpha \leftrightarrow \sigma$ we 
only need the corresponding symmetric part of the function $E$. 
This quantity is found to be
\begin{equation}
E(\alpha, \beta, \gamma, \sigma) \rightarrow -2i J_\nu \epsilon_{\eta \alpha \beta}
\{ \delta_{\alpha \sigma} \delta_{\beta \gamma} 
+(1-{\mid} V_{\gamma 1 }{\mid}^2)(1- \delta_{\alpha \sigma})
- (1-{\mid} V_{\sigma 1 }{\mid}^2) \delta_{\beta \gamma} \}  
\end{equation}
where the dummy index $\eta$ is summed from one to three.
Multiplying the above factors, in Eq.(\ref{defr}), yields
\begin{equation}
\rho R = -16 i \xi \eta v(S_l)~J_\nu X e^{2i\xi trS_\nu}
\label{rhor}
\end{equation}
where
\begin{equation}
X={1 \over 2} \sum_{\eta \alpha \beta}^{}\epsilon_{\eta \alpha \beta}
{sin(\phi_{\alpha \beta}) \over \phi_{\alpha \beta}}
\{\phi_{\alpha \beta} cos(\phi_{\alpha \beta})+2{\mid} V_{\alpha 1 }{\mid}^2
[cos(\phi_{\beta \eta }) sin(\phi_{\eta \alpha}) - \phi_{\alpha \beta} 
cos(\phi_{\alpha \beta}) ] \}
\label{defx}
\end{equation}
Note that $X$ vanishes, as
it should, if any two neutrinos would be degenerate. Furthermore,
the measurable quantities ${\mid} V_{\alpha 1 }{\mid}^2$ 
are themselves invariant functions of lepton mass matrices. They can be
extracted \cite{ceja87} from the mass matrices with help of  
projection operators $P$. In our case
we have 
\begin{equation}
{\mid} V_{\alpha 1 }{\mid}^2 = {1 \over \rho}tr (P_{\alpha} (S_\nu) {\bf \Pi } )= 
{1 \over \rho} {tr[(S_\nu -m^2_\beta )(S_\nu -m^2_\eta ) {\bf \Pi }]
\over (m^2_\alpha -m^2_\beta ) (m^2_\alpha -m^2_\eta ) }
\end{equation}
where, as before, $\xi (m^2_\alpha -m^2_\beta ) = \phi_{\alpha \beta}$. 
The identity
\begin{equation} 
sin(\phi_{\alpha \beta})cos(\phi_{\beta \eta }) sin(\phi_{\eta \alpha})
={1 \over 4} [cos(2\phi_{\alpha \beta})+cos(2\phi_{\beta \eta})
+cos(2\phi_{\eta \alpha}) -1 ]
\label{identity}
\end{equation} 
valid for $\phi_{\alpha \beta}+ \phi_{\beta \eta }+\phi_{\eta \alpha}=0$,
can be used to rewrite $X$ noting that the RHS of Eq.(\ref{identity})
has the same value for every term in $X$. 

Returning to Eq.(\ref{orderrho}), the LHS is a function of matter variables and 
the RHS contains only vacuum variables. Furthermore, the vacuum and
the first order correction terms have the same phase.
Removing common factors, we find
\begin{equation}
J^\prime_\nu~sin(\phi^\prime_{12}) 
sin(\phi^\prime_{23}) sin(\phi^\prime_{31})e^{2i\xi\rho } 
= J_{\nu} \{ sin(\phi_{12}) 
sin(\phi_{23}) sin(\phi_{31}) -i \xi \rho X + {\it O} (\rho^2) \}
\end{equation}
Thus
\begin{equation}
J^\prime_\nu~sin(\phi^\prime_{12}) 
sin(\phi^\prime_{23}) sin(\phi^\prime_{31})= J_{\nu}\ [sin(\phi_{12}) 
sin(\phi_{23}) sin(\phi_{31}) + {\it O} ( \rho^2)
\label{vacmat}
\end{equation}
Substituting the relation $J^\prime_\nu ~v(S^\prime_\nu) =J_{\nu}~v(S_\nu)$,
from Eq.(\ref{vacmat}), we find
\begin{equation}
{1 \over v(S^\prime_\nu)}sin(\phi^\prime_{12})
sin(\phi^\prime_{23}) sin(\phi^\prime_{31})= 
{1 \over v(S_\nu)}  sin(\phi_{12}) 
sin(\phi_{23}) sin(\phi_{31}) + {\it O} (\rho^2)
\end{equation}

\end{document}